\documentclass[letter]{article}

\usepackage{color,graphicx}
\usepackage{vmargin}
\usepackage{subeqnarray}
\usepackage[english]{babel}
\usepackage{amsmath,amsfonts,amssymb}

\def\c{\chi^2}
\def\s{\sigma^2}
\def\c{\chi^2}
\def\s{\sigma^2}

\newcommand{\eq}[1]{\begin{equation}#1\end{equation}}
\newcommand{\mathenv}[2]{\subparagraph{#1}{\it #2}}
\newcommand{\eqa}[1]{\begin{eqnarray}#1\end{eqnarray}}
\newcommand{\seqa}[1]{\begin{subeqnarray}#1\end{subeqnarray}}

\title{
Sequential Probability Ratio Test for the detection of a single
electron spin in the OSCAR setup\footnote{This work was supported
by the DARPA MOSAIC program under ARO contract DAAD19-02-C-0055}}
\author{Cyril Hory and Alfred O. Hero\\
System Division\\
Dept. of Electrical Engineering and Computer Science\\
University of Michigan\\
{\tt e-mail: lastname@eecs.umich.edu} }
\begin{document}

\date{}

\maketitle

 \begin{abstract}
 The MRFM device is a powerful setup for manipulating single
 electron spin in resonance in a magnetic field. However, the
 real time observation of a resonating spin is still an issue
 because of the very low SNR of the output signal. This paper
 investigates the usability and the efficiency of sequential
 detection schemes (the Sequential Probability Ratio Test) to
 decrease the required integration time, in comparison to standard
 fixed time detection schemes.
 \end{abstract}

\section{Introduction}

Magnetic Resonance Force Microscopy (MRFM) is a promising
technique for high-resolution non-destructive spatial imaging. One
of the most exciting challenge proposed for MRFM is the
observation of single spin in resonance in a magnetic field.
Sidles demonstrated the capability of the MRFM for manipulating
proton spins \cite{sid91}. The use of MRFM has since been extended
to the observation of electron spins through the use of the
OScillating Cantilever-driven Adiabatic Reversal (OSCAR) method
\cite{mam03,mik03}. Although micro-size ensembles of electron
spins have been detected \cite{zug96}, generating forces as low as
$8\times 10^{-10}$ Newton \cite{mam01}, observing a single
electron spin in resonance is still an issue because of the
weakness of the signal. For the current signal-to-noise ratio
(SNR), the required integration time for detection is too long to
allow a real-time implementation. The integration time is expected
to decrease as technological advances occur improving cantilever
sensitivity to single spins and decreasing noise sensitivity.
Improvements can also be obtained by making use of advanced signal
processing techniques. This is the focus of this paper.\\

Currently, the presence of one (or several) electron spins in
resonance is detected by standard methods of statistical
detection. A statistic of an observed sample of fixed length is
compared to a threshold. The setting of this threshold splits the
space of the statistic into two decision regions. A level of
confidence (i.e. a probability of error) is associated to each
decision regions. The probability of error decreases when the
observation time (the number of data) increases. In order to reach
an acceptable level of confidence, the observation time
is currently of the order of eight hours.\\

Sequential analysis was introduced for generic hypothesis testing
problems by Wald in 1947 \cite{wal47} to make decision with a
reduced amount of data. This is of great interest for instance in
clinical trials where ethical considerations require making
decision as soon as possible \cite{arm57}, \cite{sie85}. The most
important feature of Wald's procedure is that the number of data
required to make a decision is a random variable. The borders of
the decision regions (and thus the probabilities of error) depend
on this random variable through its expected value called the
Average Sample Number (ASN).\\
The first Sequential Probability Ratio Test (SPRT), proposed by
Wald was designed to test a simple hypothesis $H_0$ versus a
simple alternative $H_1$. Data are recorded and tested
sequentially until a condition on the likelihood ratio to accept
one of the hypotheses $H_i$ is met. The ASN of the SPRT is smaller
than the amount of data required for any fixed sample
size test to achieve the same decision error probabilities.\\

However, the SPRT presents two main drawbacks. Practically
speaking, the absence of any upper bound on the stopping time may
make the ASN higher than the actual number of data available.
Moreover, if there is a mismatch of the $H_0$ or $H_1$ models and
the data, the expected stopping time may be large and,
consequently, a sequential procedure may not improve on a fixed
sample size procedure \cite{bec60}. Much work has been done to
avoid such shortcomings. The main feature of the modified SPRT
proposed in the literature is to introduce a bound on the stopping
time \cite{and60}, \cite{arm57}. This led to a class of test
called Truncated Sequential Probability Ratio Test (TSPRT). When
performing a TSPRT, a decision is taken at a given sample size $N$
even if neither of the stopping conditions has been met before
$N$. Such modification increases the error probabilities. Another
way to deal with the uncertainty and mismatch on the hypotheses is
to take into account a priori information by means of the Bayes
formalism \cite{kie57}. Many monographs have been published since
the early book of Wald. Most of them adopt a probabilistic
approach \cite{sie85}, \cite{wet86}, by focussing on the error
probability aspect. Wijsman's approach in \cite{wij91} is slightly
different; since it is common to consider sequential procedures by
means of Brownian Motion, his analysis of sequential test (and
more generally sequential procedures) relies on elements of the
theory of diffusion which makes it an original introduction to
sequential
analysis.\\

The aim of this paper is to investigate the usability of the SPRT
in the specific problem of detecting a single electron spin in the
OSCAR setting. We focus on two sequential tests of the variance of
the observed signal, namely a $\c$ and a Fisher-F test. The main
contribution of this paper is to derive the exact expression of
the ASN of the $\c$ test and a low snr development of the ASN of
the Fisher-F test. These expressions allow a comparison to the
number of data required to perform the corresponding fixed sample
size tests. A procedure is also proposed to perform an
experimental validation of this comparison in the case of the $\c$
test.

\section{Data processing and fixed sample size detection scheme}

We address the problem of detecting an electron in resonance in
the OSCAR experiment. Before performing the detection procedure,
the data are pre-processed in order to enhance the performances of
the detector. In this section we briefly describe the OSCAR setup
and the associated signal processing.

\subsection{Data models}

\subsubsection{General outline}

 \begin{figure}[h!]
 \centering
 \input{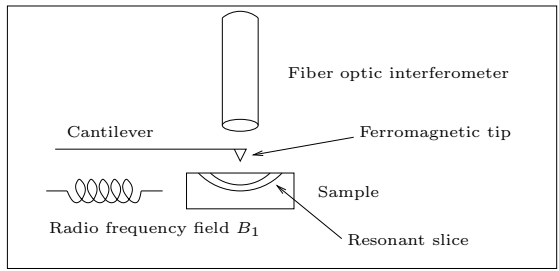}
 \caption{The OSCAR setup.}
 \label{mrfm_fig}
 \end{figure}
\begin{figure}[b!]
 \centering
 \input{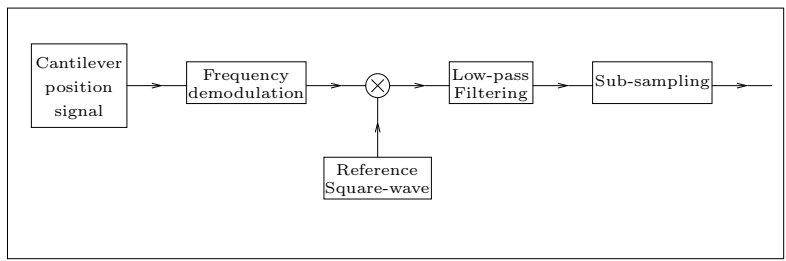}
 \caption{The pre-processing scheme.}
 \label{pre_process}
 \end{figure}

In the OSCAR experiment, presented in Figure~\ref{mrfm_fig}, a
sample is embedded in a Radio-Frequency (RF) magnetic field $B_1$.
If the RF-field frequency matches the Larmor frequency of an
electron in the sample, the electron spin is in magnetic resonance
\cite{coh77}. A cantilever with a ferromagnet on its tip is
settled close to the sample. The cantilever is forced into
mechanical oscillations at frequency $\omega_0$ which induces an
oscillating magnetic field. As a consequence, the spin polarity of
any free electron in the resonant slice of the device is forced to
reverse synchronously with the ferromagnet motion. Moreover, in
the so-called interrupted OSCAR experiment, the RF-field is turned
off every $T_{skip}$ seconds so that the spin polarity is reversed
periodically. The successive steps of the pre-processing are
presented on Figure~\ref{pre_process}.

\paragraph{Output of the interferometer.}
The spin reversal induces a slight change in the cantilever
stiffness. The electron spin can thus be detected by observing a
shift $\delta\omega_0$ in the natural frequency of the cantilever.
The motion of the cantilever is measured by a laser
interferometer. When a spin is resonating in the resonant slice,
the output of the interferometer is a frequency-modulated signal
of form:
 \eq{
 z(t)=A\cos\{\omega_0t+\int_{0}^{t}s(u)du+\phi\},
 }
where $A$ is the amplitude of the oscillation, $\phi$ is a random
phase and $s(u)$ is a square-wave of period $2T_{skip}$ and
magnitude $\delta \omega_0$. When no electron in the resonant
slice is resonating, there is no shift in the natural frequency
and $z(t)=A\cos\{\omega_0t+\phi\}$.
 \subparagraph{Spin relaxation.}
The spin can spontaneously go out of alignment with the magnetic
field. This phenomenon called spin relaxation is not fully
understood. One model for the effect of spin relaxation is that
when relaxation occurs, the polarity of the spin changes. This
causes $\lambda$ random flips per second. The relaxation
phenomenon is taken into account in the modelling of the output of
the cantilever by means of a random telegraph signal. In
continuous time, the number of random flips is considered as
following a Poisson distribution with rate $\lambda T$ where $T$
is the duration of the observation. The equivalent discrete-time
model is a $2$-state Markov chain. If the transition probabilities
$p$ between states are equal, then $p=1-T_s\lambda$ where $T_s$ is
the sampling rate. Other models include random walks on the sphere
and random walks on the interval. See \cite{mik03} for a review of
these different models.
\paragraph{Frequency demodulation and sampling.}
The square-wave is estimated by demodulating the signal with a
frequency lock-in device. Then the output of the frequency lock-in
is sampled. The frequency lock-in can be seen as a frequency
estimator whose variance induces an additive noise. Considerations
on the dynamical system describing the spin-cantilever interaction
lead to a natural alternative which makes use of an efficient
estimator of the square-wave related to the MUSIC algorithm
\cite{ssp03}.
\paragraph{Output of the correlator and filtering.}
In order to remove the deterministic square-wave, the discrete
signal is correlated to a reference square-wave of period
$2T_{skip}$ resulting in a baseband signal with the natural
frequency $\omega_0$ removed. This is the so-called {\it in phase
filtering} of the output of the frequency
estimator.\\
The signal-to-noise ratio is increased by filtering the sampled
data over the pass-band of the spin signal with a low-pass filter
defined by the recursive relation:
 \eq{
 x_n=\alpha x_{n-1}+\frac{1-\alpha}{2}(z_n+z_{n-1}),
 }
where $z_n$ is the input of the filter and $x_n$, the output. The
cut-off frequency $\omega_c$ of this recursive filter is set by
parameter $\alpha$:
 \eq{
 \alpha=\frac{1-\sin\{\omega_c\}}{\cos\{\omega_c\}}.
 }
This filtering induces a coloration of the embedding noise. In
order to consider that the assumption on independence of the data
samples is valid, the output of the filter is subsampled at the
rate $2\pi/\omega_c$.

\subsubsection{SNR estimation}
A noise-alone reference can be generated by correlating the
demodulated signal with a version of the reference square-wave
phase-shifted by $90^o$. This is the so-called {\it quadrature
filtering}. The ratio of the energies of the output of the
in-phase channel to the output of the quadrature channel provides
an estimate of the signal-to-noise ratio. Under an i.i.d. Gaussian
assumption on the sampled demodulated cantilever signal, this
ratio is by definition a Fisher-F random variable. It will be used
as the test statistic for the so-called Fisher-F test that we
describe in this paper.
 \begin{figure}[h!]
 \centering
 \input{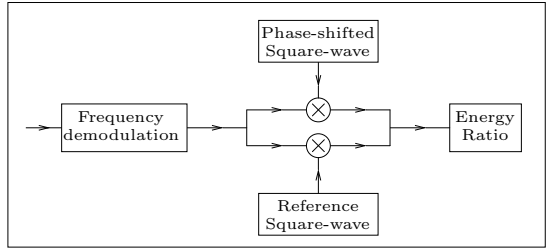}
 \caption{The SNR estimation scheme.}
 \label{snr_est}
 \end{figure}

\subsection{The energy detector.}
The single spin detection problem can be formulated as making a
decision between the two hypotheses:
 \eq{
 \left \{
 \begin{array}{ll}
 H_0: & x_{n}=\nu_{n},\\
 H_1: & x_{n}=d_{n}+\nu_{n},
 \end{array}
 \right .
 \label{hyp_start}
 }
where $d_{n}$ is a random signal with RMS amplitude $\sigma_d$ and
intensity $\lambda$ and $\nu_{n}$ is a white Gaussian noise with
zero-mean and variance $\sigma^2_{\nu}$. When the SNR
$\s_d/\sigma^2_{\nu}$ is sufficiently small we can consider $x_n$
as a Gaussian random variable with zero-mean and variance
$\sigma^2_{\nu}+\sigma^2_d$ under $H_1$. In this case the single
spin test consists in deciding:
 \eq{
 \left
 \{
 \begin{array}{ll}
 H_0: & x_{n}\sim{\cal N}(0,\sigma^2_{\nu}),\\
 H_1: & x_{n}\sim{\cal N}(0,\sigma^2_{\nu}+\sigma^2_d),
 \end{array}
 \right .
  \label{testdefint}
 }
where $X\sim Y$ means that the random variable $X$ has the same
PDF as the random variable $Y$. The detection procedure is applied
to the energy $E^{(N)}_x=\sum_{n=1}^Nx_n^2$. The energy
$E^{(N)}_x$ under both hypotheses are sums of $N$ independent
identically distributed squared Gaussian variables. They are
random variables following $\c$ distributions with $N$ degrees of
freedom and scale parameter equal to the variance of the Gaussian
variables. The hypotheses~(\ref{testdefint}) can be equivalently
formulated as:
 \eq{
 \left \{
 \begin{array}{ll}
 H_0: & E^{(N)}_x\sim\sigma^2_{\nu}\chi^2_N,\\
 H_1: & E^{(N)}_x\sim(\sigma^2_{\nu}+\s_d)\chi^2_N.
 \end{array}
 \right .
 \label{testdef}}
The Probability Density Function (PDF) of a $\s\c_{df}$ random
variable with scale parameter $\s$ and $df$ degrees of freedom is
of the form:
 \eq{
 f_{\s\c_{df}}(x)=\frac{1}{(2\s)^{df/2}\Gamma(df/2)}x^{df/2-1}e^{-x/2\s},
 \label{defchi2}
 }
where $\Gamma(u)$ is the gamma function.

\subsection{The Fisher-F test}

When $\s_{\nu}$ is unknown, the $\c$-test derived from the energy
detector can no longer be applied. An alternative detection
procedure is based on the ratio of the energies of the quadrature
and in-phase channel components. Under the hypotheses given
by~(\ref{hyp_start}), the models for the in-phase channel
component $x^i_n$ and for the quadrature component channel $x^q_n$
are:
 \eq{
 \left
 \{
 \begin{array}{llcl}
 H_0: & x^i_{n}\sim{\cal N}(0,\sigma^2_{\nu}) & \mbox{and} &x^q_{n}\sim{\cal N}(0,\sigma^2_{\nu}),\\
 H_1: & x^i_{n}\sim{\cal N}(0,\sigma^2_{\nu}+\s_d) & \mbox{and} &x^q_{n}\sim{\cal N}(0,\sigma^2_{\nu}).
 \end{array}
 \right .
 }
The detection procedure is applied to the ratio $R^{(N)}_{x}$ of
the energies:
 \eq{
 R^{(N)}_{x}=E^{(N)}_{x^i}/E^{(N)}_{x^q}.
 }
As already mentioned in the case of the energy detector, under the
Gaussian assumption on $x$ the energies are $\c$ distributed with
$N$ degrees of freedom and scale parameter equal to the variance
of the Gaussian variables. The ratio of the $\c$ variables is a
Fisher-F random variable denoted $\sigma^2{\cal F}(N,N)$ with $N$
and $N$ degrees of freedom and scale parameter $\s$ equal to the
ratio of the scale parameters of the $\c$ variables. Thus the
Fisher-F detector can be formulated as testing the two hypotheses:
 \eq{
 \left \{
 \begin{array}{ll}
 H_0: & R^{(N)}_x\sim{\cal F}(N,N),\\
 H_1: & R^{(N)}_x\sim(1+snr){\cal F}(N,N),
 \end{array}
 \right .
 \label{testfishdef}}
where the random variable $\s{\cal F}(df,df)$ has a Fisher-F PDF
of the form:
 \eq{
 f_{\sigma^2{\cal
 F}(df,df)}(x)=\frac{1}{\sigma^2}\frac{\Gamma(df)}{\Gamma(df/2)^2}
 \frac{(x/\sigma^2)^{df/2-1}}{(1+x/\sigma^2)^{df}}.
 \label{deffisher}
 }

\subsection{Likelihood Ratio Tests}

\subsubsection{Generality}
Given a random variable ${\bf X}$ having a PDF $f_{\theta}({\bf
x})$ defined by the parameter $\theta$, the likelihood function
$l(\theta;{\bf x})$ of a sample ${\bf x}$ of ${\bf X}$ is defined
as $l(\theta;{\bf x})=f_{\theta}({\bf x})$. The Likelihood Ratio
Test (LRT) consists in testing simple hypothesis $H_0$ under which
$X$ follows $f_0({\bf x})=f_{\theta_0}({\bf x})$ versus the simple
alternative $H_1$ under which $X$ follows $f_1({\bf
x})=f_{\theta_1}({\bf x})$ by setting a threshold $\tau$ on the
likelihood ratio statistic:
 \eq{
 \lambda({\bf x})=\frac{f_1({\bf x})}{f_0({\bf x})}.
 \label{like_ratio}
 }
If $\lambda({\bf x})>\tau$ hypothesis $H_0$ is rejected and
vice-versa. The notation:
 \eq{
 \begin{array}{ccc}
 &H_1&\\
 \lambda({\bf x})&\gtrless&\tau,\\
 &H_0&
 \end{array}
 }
is often adopted to evoke the LRT.
\paragraph{Error probabilities of the test}
The LRT is intended at taking a decision given a finite sample of
a random variable. The performance of a test is evaluated by the
decision error probabilities. The {\it probability of miss}
$P_{mis}$ (also called the probability of error of first kind) is
the probability of deciding $H_0$ when $H_1$ is true and the {\it
probability of false alarm} $P_{fa}$ (also called the probability
of error of second kind) is the probability of deciding $H_1$ when
$H_0$ is true. The threshold $\tau$ determines the error
probabilities. An ideal test would require to choose $\tau$ such
that both error probabilities tend to zero. Unfortunately the
probability of miss increases when the probability of false alarm
decreases. The user has to decide which one of the errors is
preferable. However, the LRT is often referred as the optimal test
in the sense that for a given $P_{fa}$, the LRT provides the
smallest $P_{mis}$.

\paragraph{Decision regions}
The LRT splits the sample space into two exclusive regions of
acceptance of $H_0$ or $H_1$. For fixed sample size $N$, the
boundary between the decision regions depends on the probabilities
of error through the setting of $\tau$ and $N$. Thus, given a
sample size $N$ and a threshold $\tau$, $P_{mis}$ and $P_{fa}$ are
uniquely defined.

\subsubsection{The Energy detector\label{test_chi2}}

For the LRT involving the likelihood function~(\ref{defchi2}) of
the energy $E^{(N)}_x$, the test statistic~(\ref{like_ratio})
takes the form:
 \eq{
 \lambda^{(N)}({\bf x})=\left ( \frac{1}{1+snr}\right )^{N/2}\mbox{exp}\{\frac{1}{2\sigma_{\nu}^2}\frac{snr}{1+snr}E^{(N)}_x\},
 }
where for sake of simplicity in the notations, the dependence on
$E^{(N)}_x$ is replaced by a dependence on ${\bf
x}=x_1,\ldots,x_N$ and $snr=\frac{\s_d}{\s_{\nu}}$. It is often
more convenient to test the log-likelihood $\Lambda^{(N)}({\bf
x})$ of the data:
 \eq{
 \Lambda^{(N)}({\bf x})=\log\{\lambda^{(N)}({\bf x})\}=\frac{snr}{1+snr}\frac{1}{2\sigma^2_{\nu}}E^{(N)}_x-\frac{N}{2}\log\{1+snr\}.
 \label{loglike}
  }
One can see that the log-likelihood ratio statistic depends on the
energy $E^{(N)}_x$ through parameters $snr$ and $\sigma^2_{\nu}$
which are common to both hypotheses. Hence, when testing two
normal distributions with zero mean and different known variances,
the energy detector is equivalent to the LRT.

\subsubsection{The Fisher-F Ratio detector}

The log-likelihood ratio of the energy ratio $R^{(N)}_x$ is
defined from expression~(\ref{deffisher}) of the Fisher-F PDF by:
 \eq{
 \Lambda^{(N)}(x)=\frac{N}{2}\log\{1+snr\}+N\log\{1+x\}-N\log\{1+snr+x\}.
 \label{loglikefish}
 }
Unlike the log-likelihood ratio~(\ref{loglike}) involved in the
energy detector, this test statistic depends only on the
signal-to-noise ratio $snr$. Knowledge on the noise variance is
not required to implement the Fisher-F test. This is the reason
why this test is currently preferred in the OSCAR experiment.

\subsubsection{Required sample number}

Expressions~(\ref{testdef})~and~(\ref{testfishdef}) show that the
parameter of interest for discriminating between the hypotheses is
the scale parameter $\s_i$. In both cases, the scale parameters
are such that $\s_1=(1+snr)\s_0$. The cumulative distribution
functions $F_{i}(\tau)=\int_{0}^{\tau}f_{i}(u)\mbox{d}u$ of the
$\c$ and the Fisher-F random variables for $i=0,1$ satisfy the
relation:
 \eq{
 F_{1}(\tau)=F_{0}(\frac{\tau}{1+snr}).
 }
The probability of miss can thus be expressed as a function of the
probability of detection, parameterized by $snr$, and the sample
size $N$. The required number of samples can be evaluated for a
given test with a given strength $(P_{fa},P_{mis})=(\alpha,\beta)$
by constraining $P_{fa}$, and choosing $N$ such that $P_{mis}$ is
reached.

\section{Sequential Detection Procedures}
Unlike fixed sample size procedures, the sample size required in a
sequential procedure is a random number called the sample number.
The Average Sample Number (ASN) can be dramatically smaller than
the corresponding sample size $N$ required to perform a fixed
sample size test of at least same strength $(\alpha,\beta)$
\cite{sie85,wal47}. The Relative Sample Efficiency (RSE) of a
sequential procedure with respect to a fixed sample size procedure
is the ratio $N/\mbox{ASN}$.
\subsection{Sequential Probability Ratio Test}
A fixed sample size test splits the sample space into two decision
regions. A sequential procedure splits the sample space into three
regions; the region of acceptance of $H_0$, the region of
acceptance of $H_1$, and the continuation region where the
decision is postponed and another sample is acquired. More
specifically, suppose a decision has to be taken from the
likelihood ratio statistic $\lambda^{(n)}$ computed from the $n$
first available samples $x_1,x_2,\ldots,x_n$. Two constants $A$
and $B$ are chosen such that $A>B$ and the SPRT is defined as
follows:
 \eq{
 \left \{
 \begin{array}{ll}
 \lambda^{(n)} \leq B: & \mbox{accept } H_0,\\
 B < \lambda^{(n)} < A: & \mbox{postpone the decision},\\
 \lambda^{(n)} \geq A: & \mbox{accept } H_1
 \end{array}
 \right .
 }
The thresholds $A$ and $B$ define the boundaries between the three
decision regions. The stopping time or sample number $N_s$ is
defined by:
 \eq{
 N_s = \min\{N_0,N_1\},
 }
where $N_0$ (resp. $N_1$) is the first time the likelihood ratio
statistic crosses the boundary $A$ (resp. $B$). The probability
that such a test terminates is one. Wald and Wolfowitz have shown
that among all the sequential test, the SPRT provides the smallest
ASN under both hypotheses \cite{wal48}. For a desired strength
$(\alpha,\beta)$ of test, the boundaries are set by the Wald's
approximations \cite{joh61,wal47}:
 \seqa{
 A & = & \frac{1-\beta}{\alpha}\\
 B & = &\frac{\beta}{1-\alpha}.
 \label{approx}
 }
These approximations hold for small error probabilities, typically
smaller than $0.05$.
\paragraph{Operating Characteristic and Average Sample Number function}
The design of the SPRT assumes that the parameters $\theta_i$
defining the hypotheses are known. The behavior of the test
strongly depends on this assumption. In particular, the ASN can
dramatically increase if the true parameter $\theta$ does not
match the hypotheses. The Operating Characteristic $L(\theta)$ is
the probability of accepting $H_0$ when the true parameter is
$\theta$. In particular, $L(\theta_0)=1-P_{fa}$ and
$L(\theta_1)=P_{mis}$. The operating characteristic is an
efficient tool for evaluating the performances of the test under a
model mismatch.\\
Wald shows that the operating characteristic of the SPRT can be
approximated by:
 \eq{
 L(\theta)\approx\frac{A^{h}-1}{A^{h}-B^{h}},
 \label{oc}
 }
where $h$ is solution of the integral equation:
 \eq{
 \int_{-\infty}^{+\infty}\left (\frac{f(x,\theta_1)}{f(x,\theta_0)}\right )^{h}f(x,\theta)dx=1.
 \label{eq_int}
 }
One can see from this equation that $h$ depends on the true
parameter $\theta$. For instance the solutions in the cases
$\theta=\theta_0$ and $\theta=\theta_1$ can be computed by noting
that $f(x,\theta)$ is a PDF:
 \eq{
 \left \{
 \begin{array}{lcc}
 h=1&\mbox{if}&\theta=\theta_0,\\
 h=-1&\mbox{if}&\theta=\theta_1.
 \end{array}
 \right .
 }
 In
general, $h$ cannot be evaluated explicitly for every $\theta$ and
one has to approximate it numerically. A possible approach
suggested by Wald consists of solving equation~(\ref{eq_int}) by
finding the value of $\theta$ for which
the equation is verified by a given value of $h$ \cite{wal47}.\\
Wald makes use of the operating characteristic~(\ref{oc}) and of
the approximations~(\ref{approx}) to derive an approximation to
the ASN $E\{N|\theta\}$ required to stop the SPRT when the true
parameter is $\theta$:
 \eq{
 E\{N|\theta\}=\frac{L(\theta)\log\{B\}+(1-L(\theta))\log\{A\}}{E\{\Lambda(x)|\theta\}},
 \label{approx_asn}
 }
where $\Lambda(x)=\Lambda^{(1)}(x)$. Unlike the operating
characteristic, the ASN depends on the model through the expected
value of the log-likelihood ratio.

\subsection{The Approximation of the Log-Likelihood Ratio Statistic as a Brownian Motion\label{BM}}

Most of the properties of the SPRT are model-independent. In
particular, the statistical behavior of the SPRT can be studied by
approximating the sequential log-likelihood ratio statistic as a
Brownian Motion \cite{sie85}.

\subsubsection{The Brownian Motion}

\mathenv{Definition of a Brownian Motion.}{ A Brownian Motion
(also called a Wiener process) $W(t)\sim {\cal B}{\cal
M}(\mu,\s)$, $0\leqslant t < \infty$ with drift $\mu$ and variance
$\s$ is a random process such that:
 \begin{itemize}
 \item{$W(0)=0$;}
 \item{$W(t)-W(s)\sim {\cal N}(\mu(t-s),\s(t-s))$, for all $0\leqslant s <t
 <\infty$;}
 \item{for all $0\leqslant s_1 <t_1<s_2<t_2<\infty$, $W(t)-W(s)\sim {\cal N}(\mu(t-s),\s(t-s))$, the random variables
 $W(t_1)-W(s_1)$ and $W(t_2)-W(s_2)$ are independent;}
 \item{$W(t)$, $0\leqslant t <\infty$ is a continuous function of $t$.}
 \end{itemize}
 }
If the mean and variance of $\Lambda^{(N)}({\bf x})$ are linear
functions of the sample size $N$, the log-likelihood ratio can be
considered as the sampled Brownian Motion
$W(t)=W(NT_s)=\Lambda^{(N)}({\bf x})\sim{\cal B}{\cal M}(\mu,\s)$.
Then by equating the first and second moments of $\Lambda^{(N)}$
under $H_0$ and $H_1$, one can compute the drift $\mu_i$ and the
variance $\s_i$ under hypothesis $H_i$. The test can then be
formulated under a new model:
 \eq{
 \left
 \{
 \begin{array}{ll}
 H_0: & W(t)\sim{\cal B}{\cal M}(\mu_0,\s_0),\\
 H_1: & W(t)\sim{\cal B}{\cal M}(\mu_1,\s_1),
 \end{array}
 \right .
 \label{hyp_bm}
 }
where $\mu_i$ and $\sigma_i$
are parameters characterizing the performance of the test procedure.\\
The drifts and variances are derived from the first and second
order moments of the test statistic. Thus the approximation of the
sequential likelihood ratio statistic as a sampled version of a
Brownian Motion is valid up to the second order moment. A drawback
of such an approximation is that the skewness of the distribution
is not taken into account. This leads to an overestimation of the
expected value of the stopping time in many situation. This
phenomenon is known as the {\it overshooting} \cite{wij91}. The
overshooting phenomenon has been evaluated numerically for some
tests \cite{sie85}.

\subsubsection{Truncated SPRT and prediction of the sample number} 
\begin{figure}[t!]
 \centering
 \includegraphics[width=0.48\linewidth,height=0.4\linewidth]{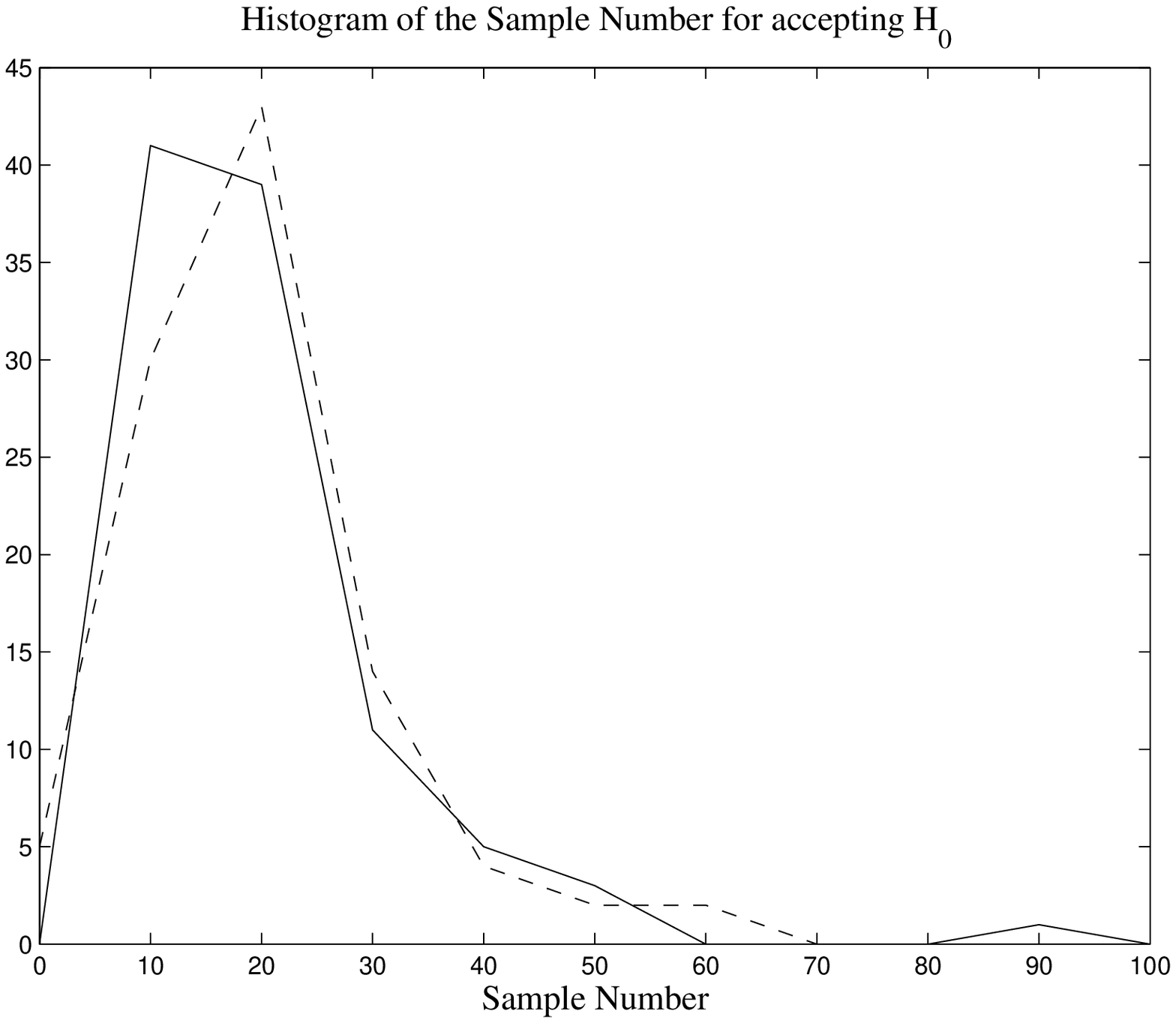}
 \includegraphics[width=0.48\linewidth,height=0.4\linewidth]{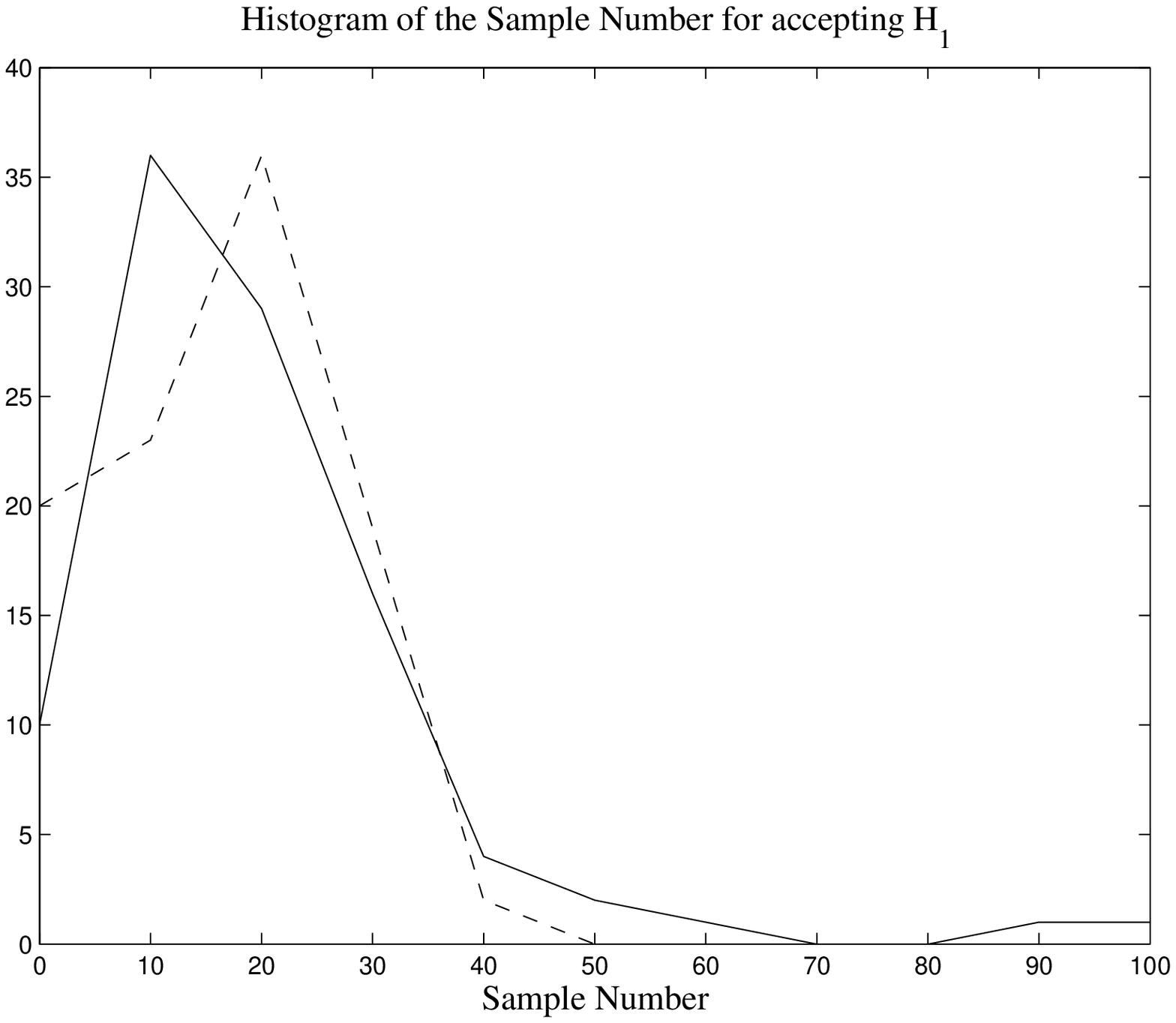}
 \caption{Experimental evaluation of the accuracy of the prediction of the required sample number. The histograms of the prediction (dashed line)
 are superimposed to the histograms of the actual sample number (plain lines).\label{fig_est}}
 \end{figure}

In many
applications, the number of samples available is limited. After a
given time, a decision has to be made even though the sample still
spans the neutral region. Define the truncated Sequential
Probability Ratio Test (TSPRT) with stopping rules:
 \eq{
 \left \{
 \begin{array}{rcl}
 T_0 &=& \min\{ \inf \{n: \Lambda^{(n)} \leqslant \log\{A\}\}, N\},\\
 T_1 &=& \min\{ \inf \{n: \Lambda^{(n)} \geqslant \log\{B\}\}, N\},
 \end{array}
 \right .
 }
where $N$ is the maximum practicable sample size. Under both
hypotheses, if $T_i<N$ then the estimated stopping time is
$N_i=T_i$, else, thanks to the definition of Brownian Motion, one
can predict what should be the stopping time if additional samples
were to be taken. Indeed, $N_i$ is such that
 \eq{
 W(N_iT_s)-W(NT_s)\sim{\cal N}(\mu(N_i-N)T_s,\s (N_i-N)T_s).
 }
By noting that $W(NT_s)=\Lambda^{(N)}$ and $W(N_iT_s)=\log\{A\}$
under $H_0$ and $W(N_iT_s)=\log\{B\}$ under $H_1$ one can express:
 \eq{
 \left
 \{
 \begin{array}{rcl}
 N_0 &=& N+\frac{1}{\mu_0}(\log\{A\}-\Lambda^{(N)}),\\
 N_1 &=& N+\frac{1}{\mu_1}(\log\{B\}-\Lambda^{(N)}).
 \end{array}
 \right .
 }
On Figure~\ref{fig_est} are presented the histograms of the
estimated ASN $E_0\{N_0\}$ and $E_1\{N_1\}$ for lengths $N=1000$
and $N=5000$ for the SPRT based on the energy statistic described
in next section. One hundred trials have been performed. The
strength of the test is $P_{fa}=_{mis}=0.02$. The signal-to-noise
ratio is $-20dB$ before filtering. It has been chosen such that
for each trial, $1000<N_0,N_1<5000$ with high probability. Thus
for each trial, when $N=1000$ (dashed lines) the sample number has
to be predicted and when $N=5000$ (plain line), it can be
observed. The tail of the true histogram is heavier than for the
predicted histogram. Indeed, the prediction scheme is based on the
approximation of the likelihood ratio as a Brownian Motion. The
distribution of the predicted sample number is closer to a
Gaussian distribution. The consequence is a reduction of the
skewness. Thus like for the approximation~(\ref{approx_asn})
proposed by Wald, the ASN is slightly under-estimated by
approximating the likelihood ratio sequence as a Brownian motion.

\subsection{SPRT based on the energy statistic}
\begin{figure}[t!]
 \centering
 \includegraphics[width=0.48\linewidth,height=0.4\linewidth]{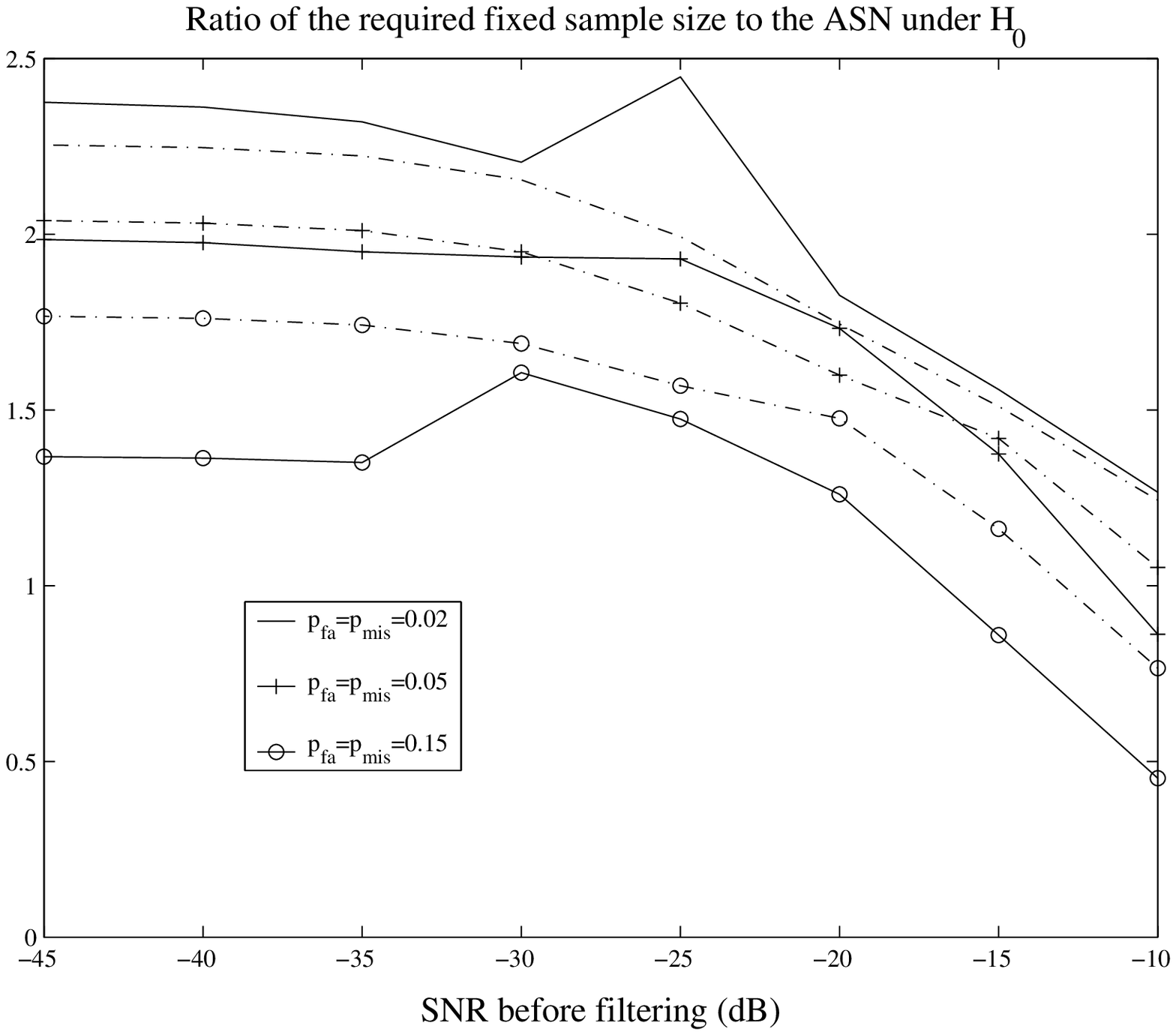}
 \includegraphics[width=0.48\linewidth,height=0.4\linewidth]{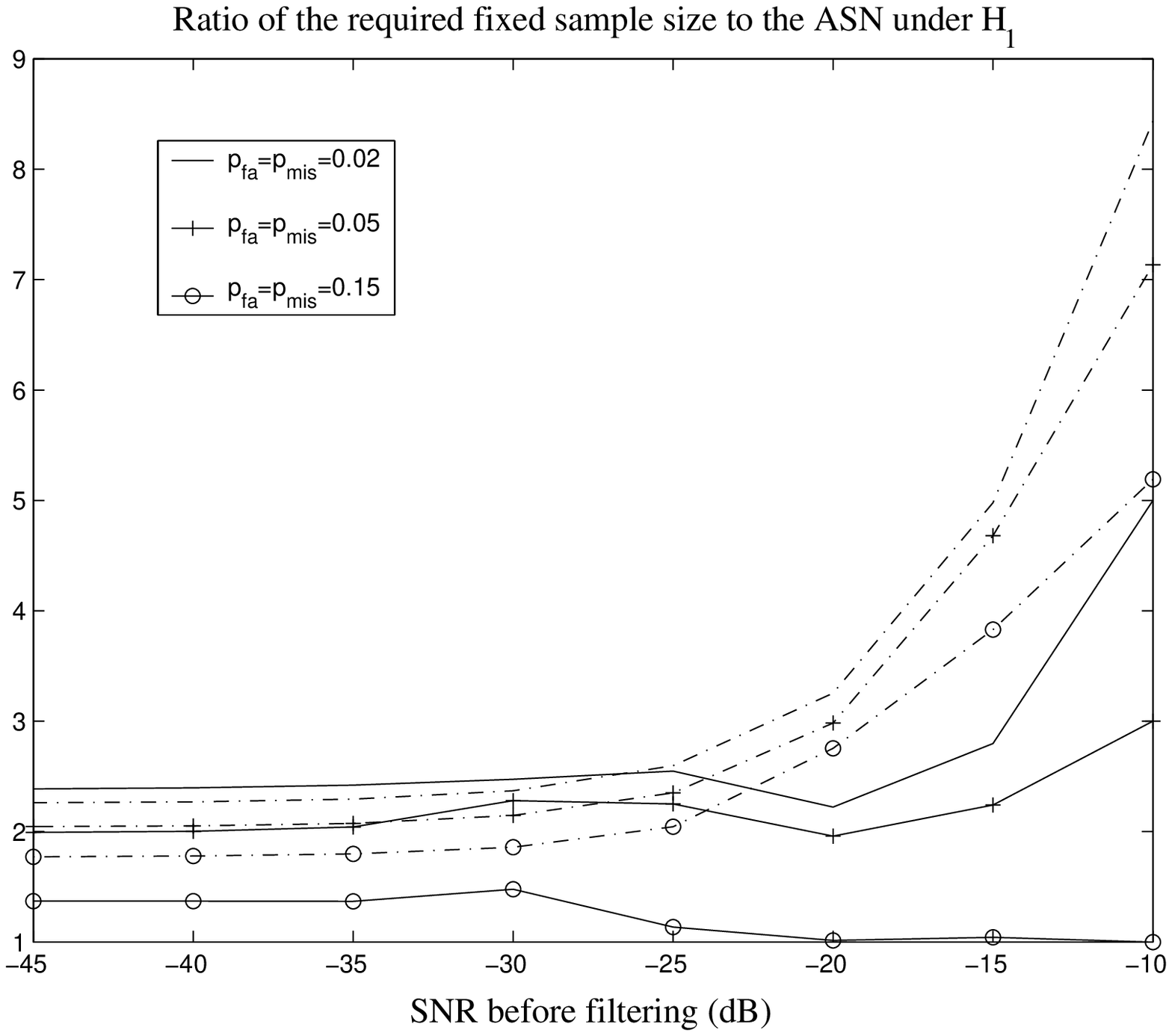}
 \caption{Evaluation of the $\c$ test: RSE of the energy detector
 under $H_0$ (left) and $H_1$ (right). The dashed-dotted lines are the theoretical
 RSE.\label{fig_snr}}
 \end{figure}

In the case of $\chi^2$ (or Gamma) distributions, Bartholomew
\cite{bar56} and Phatarfod \cite{pha71} have proposed specific
formulations of the procedure for application of a sequential test
to analysis of the arrival time of a random event. These works
concern distributions which differ under the null hypothesis and
the alternative by the number of degrees freedom. We are concerned
with distributions which present the
same number of degrees of freedom, namely, the sample number.\\
The expected value and variance of a $\s\c_N$ random variable are:
 \eqa{
 E\{\s\c_N\}=N\s.\\
 Var\{\s\c_N\}=2N(\s)^2,
 }
so, under the hypotheses~(\ref{testdef}), the test
statistic~(\ref{loglike}) can be approximated by the following
Brownian motion:
 \eq{
 \left
 \{
 \begin{array}{ll}
 H_0: & W(t)\sim{\cal B}{\cal M}(\frac{snr}{2(1+snr)}-\frac{1}{2}\log\{1+snr\},\frac{snr^2}{2(1+snr)^2}),\\
 H_1: & W(t)\sim{\cal B}{\cal M}(\frac{snr}{2}-\frac{1}{2}\log\{1+snr\},\frac{1}{2(1+snr)^2}),
 \end{array}
 \right .
 }
Figure~\ref{fig_snr} displays the ratio of the ASN to the number
of sample required for the fixed sample LRT with same error
probabilities. The ASN have been estimated as the average of the
sample numbers computed from $100$ trials of $20000$ points. After
filtering the sub-sampling reduces the sample size to $N=250$.
When the test failed to stop before $N$, the prediction procedure
described in section~\ref{BM} was applied. The number of samples
required for the fixed sample size test has been computed
numerically from the known expression of the error probabilities
as described in section~\ref{test_chi2}.\\
At higher SNR, the number of samples required for both the
sequential test and the standard test are of the order of the
unit, so the computed ratios cannot be considered as reliable. At
low SNR, the RSE is around two.
The RSE increases when the error probabilities decrease. This
makes the SPRT particularly attractive when the level of
confidence put on
the decision is to be high.\\
The ratios predicted from Wald's approximation~(\ref{approx_asn})
of the ASN are superimposed to the experimental ratios. At low
error probabilities, one can see that Wald's expression
over-estimates the true ASN. This phenomenon is associated with
the skewness of the PDF (highlighted by Figure~\ref{fig_est})
\cite{sie85, wet86}. It is of interest to notice that when the
error probabilities increase (see the case $P_{fa}=P_{mis}=0.15$),
approximation ~(\ref{approx_asn}) no longer holds and the ASN
tends to be under-estimated. However, the improvement is still
significant. Above this value, there is no gain in using a
sequential test.
\subsection{SPRT based on the Fisher-F statistic}
\begin{figure}[t!]
 \centering
 \includegraphics[width=0.48\linewidth,height=0.4\linewidth]{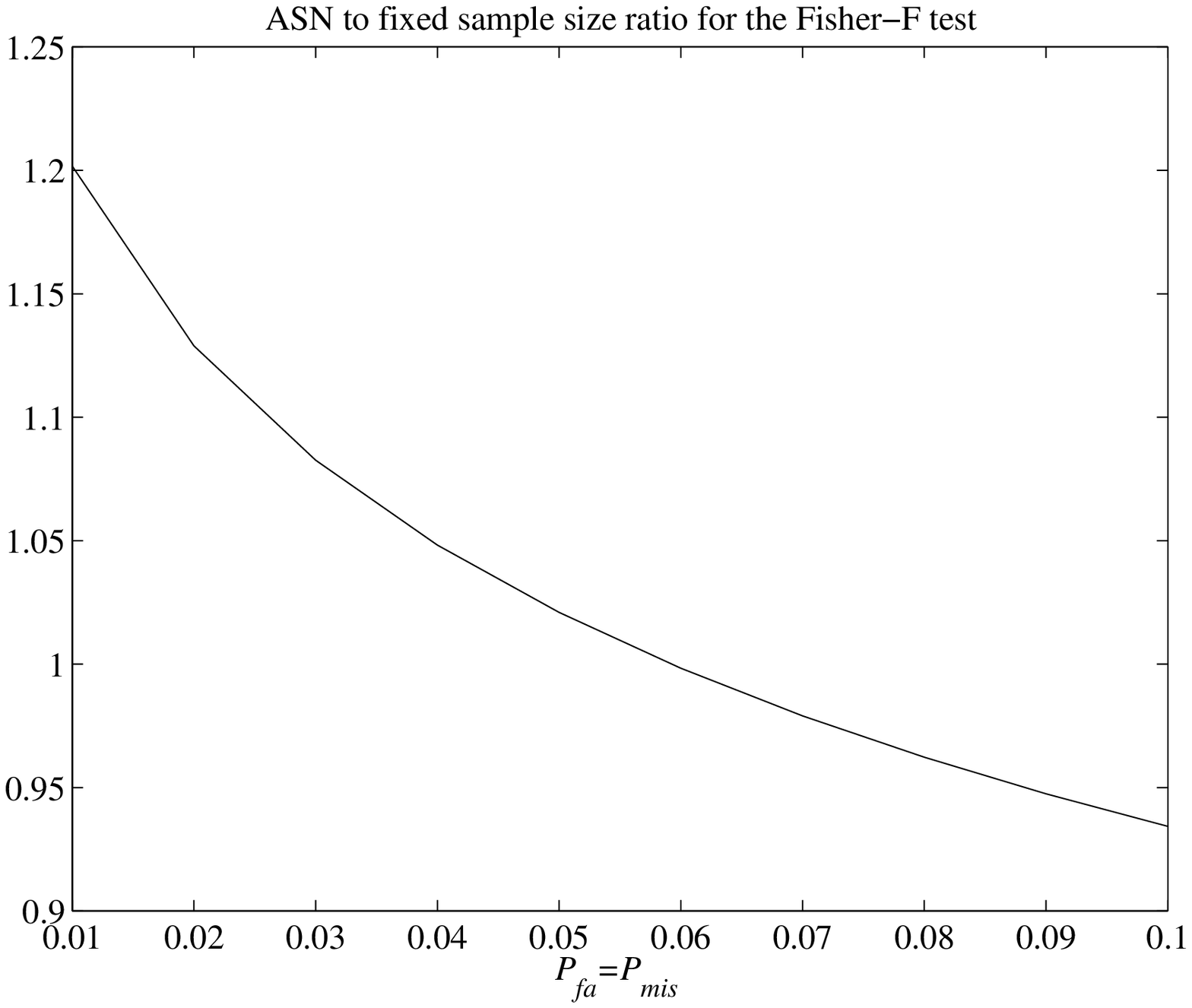}
 \includegraphics[width=0.48\linewidth,height=0.4\linewidth]{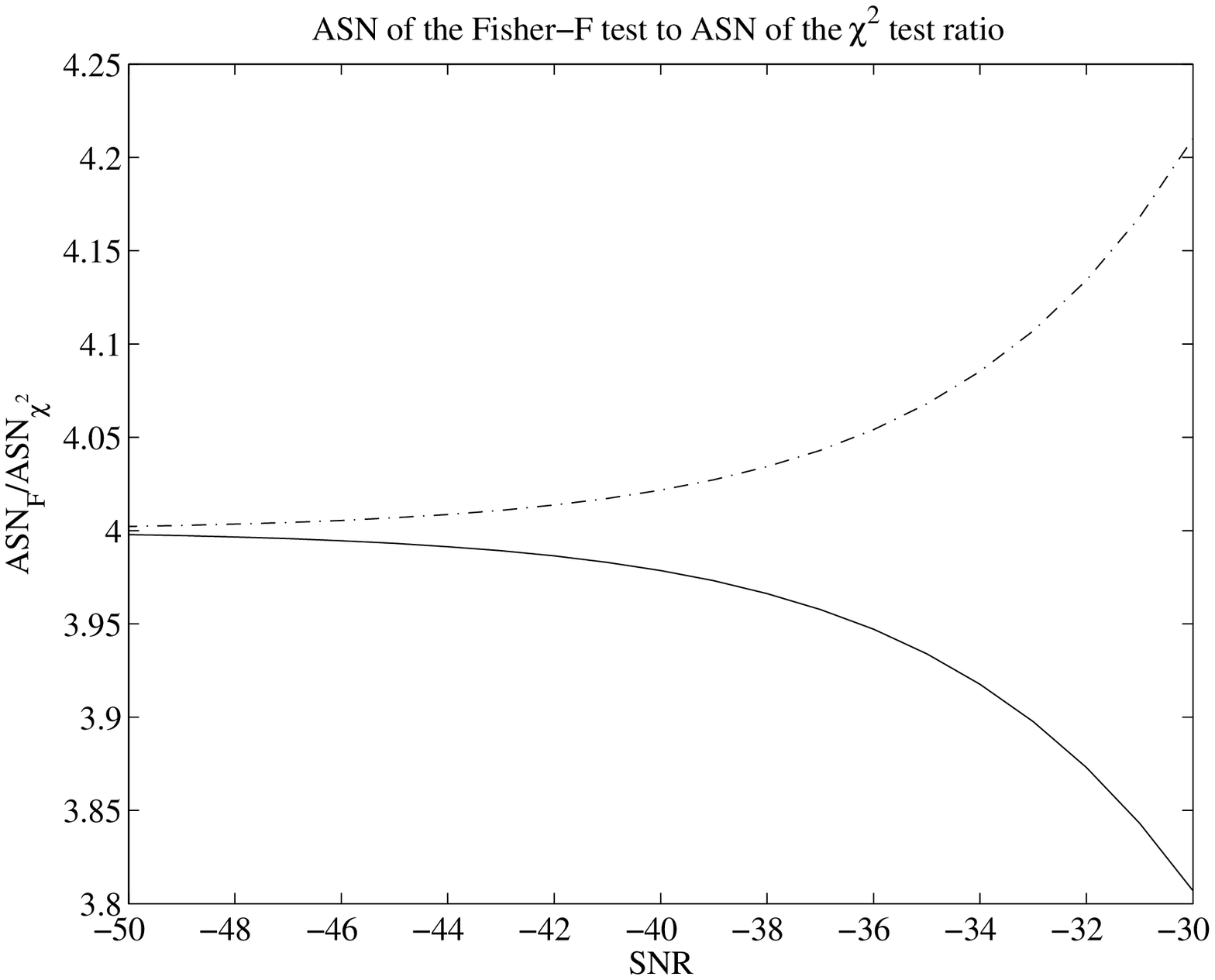}
 \caption{Evaluation of the Fisher-F test: RSE of the Fisher-F test
 under $H_1$ for various error probabilities (left). The snr before filtering is $-30dB$.
 Ratio of the ASN of the Fisher-F test to the ASN of the $\c$ test (right) uner $H_0$
 (dashed line) and under $H_1$ (plain line).\label{fig_pfa}}
 \end{figure}

Jackson \cite{jac61} and Jennison \cite{jen91} have derived exact
values of the operating curve and ASN for the sequential Fisher
test when the parameter of interest is the mean of a normal
population. Like Bartholomew in \cite{bar56} and Phatarfod in
\cite{pha71}, Jennison makes use of Cox's theorem to transform the
observation and apply the test to a new statistic which is
independent on the number of degrees of freedom. Such approach
cannot be adopted in our case as the parameter of interest is the
variance of the normal population.\\
An exact expression of the expected value of the
log-likelihood~(\ref{loglikefish}) has yet to be derived. We
propose in the Appendix a derivation of an approximation that
holds at low snr. One can see with expressions~(\ref{exp_final})
that the expected value of the log-likelihood is not a linear
function of the number of data. Thus, the approximation as a
Brownian Motion is not valid for the Fisher-F test statistic.
Consequently, the sample size prediction procedure does not hold
and the experimental evaluation of the ASN cannot be performed. On
Figure~\ref{fig_pfa} is presented the theoretical ratio of the ASN
under $H_1$ to the number of samples required for the fixed sample
size test of same strength at low snr ($-30dB$ before
filtering).\\
When the error probabilities are smaller than $0.06$, the SPRT
significantly reduces the required sample size. A strength of
$\alpha=\beta=0.02$ can be achieved by a fixed sample size test if
the sample size is of the order of $10^7$. The use of the SPRT
allows a reduction of the order of $10^6$ data samples. When the
error probabilities are above $0.06$ the ratio is smaller than
$1$, which is due to the fact that the Wald's approximation is not
reliable for large error probabilities.\\
The right-hand side of Figure~\ref{fig_pfa} displays the ASN of
the Fisher-F test to the ASN of the $\c$ test ratio computed from
expression~(\ref{approx_asn}) and approximation~(\ref{exp_final})
under $H_0$ and $H_1$. As far as Wald's approximation holds, one
can see from expression~(\ref{approx_asn}) that this ratio is
independent of the error probabilities. At low signal-to-noise
ratio, the ASN of the Fisher-F test is $4$ times the ASN of the
$\c$ test.
\section{Conclusion}

The advantages of Sequential Probability Ratio Tests (SPRT) for
detection in a $\c$ and Fisher-F model has been investigated.
Evaluation of the relative sample efficiency of the SPRT with
respect to the corresponding fixed sample size test has been
computed on simulation data.\\
For these test, it was experimentally verified that a gain of $50
\%$ in the required integration time can be expected at low SNR
(below $-30dB$). The appealing feature of the proposed model is
that the test statistic can be written in a closed form which
avoid making use of approximations.\\
A similar study has been performed to evaluate the efficiency of a
Fisher-F type test based on the ratio of the energies of the
in-phase and quadrature channel outputs. This test requires only
the knowledge of the signal-to-noise ratio and can be performed
whatever the noise variance. The consequence is an increase in the
ASN of he Fisher-F test in comparison to the $\c$ test. At low
error probabilities, the ASN is still smaller than the number of
samples required to perform
a fixed sample size Fisher-F test of same strength.\\
Another direction required for application to the OSCAR experiment
is the evaluation of SPRT performances for unknown bandwidth of
the spin signal. This will allow SPRT to be developed for the
filter bank implementation of the OSCAR experimental apparatus.

\appendix

\part*{Appendix}

\section*{Expected value of the statistic of the Fisher test}

The log-likelihood ratio of the Fisher-F test is of form:
 \eq{
 \Lambda^{(N)}(x)=\frac{N}{2}\log\{1+snr\}+N\log\{1+x\}-N\log\{1+snr+x\}.
 }
The expansion into Taylor series of this function around $snr=0$
provides the approximation for small $snr$:
 \eq{
 \Lambda^{(N)}(x)=\frac{N}{2}\log\{1+snr\}+N\sum_{m=1}^M(-1)^m\frac{snr^m}{m(1+x)^m}+o(M).
 }
Thus the expected values of the likelihood ratio under both
hypotheses involves the terms $E_i\{1/(1+x)^m\}$.

\paragraph{Computation of $E_0\{\frac{1}{(1+x)^m}\}$.}
Under $H_0$ these terms are written:
 \eqa{
 E_0\{\frac{1}{(1+x)^m}\} & = &
 \int_0^{+\infty}\frac{1}{(1+x)^m}\frac{\Gamma(N)}{\Gamma(N/2)^2}\frac{x^{N/2-1}}{(1+x)^N}dx,\\
 & = &
 \frac{\Gamma(N)}{\Gamma(N/2)^2}\int_0^{+\infty}\frac{1}{(1+x)^m}\frac{x^{N/2-1}}{(1+x)^N}dx.
 }
After the change of variable $y=\log(1+x)$, this term takes the
form:
 \eqa{
 E_0\{\frac{1}{(1+x)^m}\} & = &
 \frac{\Gamma(N)}{\Gamma(N/2)^2}\int_0^{+\infty}\mbox{e}^{-(N+m-1)y}(\mbox{e}^{y}-1)^{N/2-1}dy,\nonumber\\
 & = &\frac{\Gamma(N)}{\Gamma(N/2)^2}A_m(N,1),
 \label{exp_in}
 }
where the integral $A_m(N,b)$ is defined by:
 \eq{
 A_m(N,b)=\int_0^{+\infty}\mbox{e}^{-(N+m-b)y}(\mbox{e}^{y}-1)^{N/2-b}dy.
 }
An integration by parts involving the functions:
 \eqa{
 f(y) & = & \mbox{e}^{-(N+m-b+1)y},\nonumber\\
 g(y) & = & \frac{1}{N/2-b+1}(\mbox{e}^y-1)^{n/2-b+1},\nonumber
 }
show that under the condition $b\leq n/2$, $A_{m}(n,b)$ satisfies
the recursive equation:
 \eq{
 A_{m}(N,b)=\alpha_m(N,b)A_{m}(N,b-1),
 }
where $\alpha(N,b)=\frac{N+m-b+1}{N/2-b+1}$. This equation leads
to the relation:
 \eqa{
 \forall
 b,c&A_m(N,c)=&\frac{A_m(N,b)}{\prod_{k=c+1}^b\alpha_m(N,k)}.
 \label{rec_int}
 }
By noting that for the case $b=N/2$, $A_m(N,N/2)=2/(N+2m)$,
expression~(\ref{rec_int}) takes the form:
 \eqa{
 \forall c&A_{m}(N,c)=&\frac{2/N}{\prod_{k=c+1}^{N/2}\alpha(N,k)}.
 \label{a0nc}
 }
This equality holds for even values of $N$. It is asymptotically
true for the odd values of $N$. The denominator can be expressed
in terms of Gamma functions:
 \[
 \prod_{k=c+1}^{N/2}\alpha(N,k)=\frac{\Gamma(N+m-c+1)}{\Gamma(N/2-c+1)\Gamma(N/2+m+1)},
 \]
and the integral $A_m(N,b)$ takes the form:
 \eq{
 A_m(N,b)=\frac{\Gamma(N/2-b+1)\Gamma(N/2+m)}{\Gamma(N+m-b+1)}.
 }
By injecting this expression into~(\ref{exp_in}), the expected
value of $1/(1+x)^m$ under $H_0$ is finally written:
 \eq{
 E_0\{\frac{1}{(1+x)^m}\}=\frac{\Gamma(N)\Gamma(N/2+m)}{\Gamma(N/2)\Gamma(N+m)}.
 }

\paragraph{Computation of $E_1\{\frac{1}{(1+x)^m}\}$.}

The expected value of $1/(1+x)^m$ under $H_1$ is of form:
 \eq{
 E_1\{\frac{1}{(1+x)^m}\}=\frac{\Gamma(N)}{\Gamma(N/2)^2}
 \int_0^{+\infty}\frac{1}{(1+x)^m}\frac{1}{1+snr}\frac{(x/(1+snr)^{N/2-1}}{(1+x/(1+snr))^N}dx.
 }
After the change of variable $y=x/(1+snr)$ this expression takes
the form:
 \eq{
 E_1\{\frac{1}{(1+x)^m}\}=\frac{\Gamma(N)}{\Gamma(N/2)^2}\int_0^{+\infty}h(y,snr)\frac{y^{N/2-1}}{(1+y)^N}dx.
 }
where $h(y,snr)=\frac{1}{(1+y(1+snr))^m}$. An expansion of
$h(y,snr)$ into Taylor series around $snr=0$:
 \eq{
 h(y,snr)=\sum_{k=0}^{K}(-1)^k\frac{\Gamma(m+k)}{\Gamma(m)}\frac{(snry)^k}{k!(1+y)^{m+k}}+o(K),
 }
leads to the expression:
 \eq{
 E_1\{\frac{1}{(1+x)^m}\}\approx\frac{\Gamma(N)}{\Gamma(N/2)^2}
 \sum_{k=0}^{K}(-1)^k\frac{\Gamma(m+k)}{\Gamma(m)}\frac{snr^k}{k!}\int_0^{+\infty}
 \frac{y^k}{(1+y)^{m+k}}
 \frac{y^{N/2-1}}{(1+y)^N}dy,
 }
where one can recognize the integrals $A_m(N,1-k)$ previously
defined. Finally, the expected value of $1/(1+x)^m$ under $H_1$
takes the form:
 \eq{
 E_1\{\frac{1}{(1+x)^m}\}\approx\frac{\Gamma(N)}{\Gamma(N/2)^2}
 \sum_{k=0}^{K}(-1)^k\frac{\Gamma(m+k)}{\Gamma(m)}\frac{snr^k}{k!}A_m(N,1-k),
 }
The expected values of the likelihood ratio statistic take finally
the forms:
 \eqa{
 E_0\{\Lambda^{(N)}(x)\}&\approx&\frac{N}{2}\log\{1+snr\}+N\frac{\Gamma(N)}{\Gamma(N/2)^2}\sum_{m=1}^M(-1)^m\frac{snr^m}{m}A_m(N,1),\\
 E_1\{\Lambda^{(N)}(x)\}&\approx&\frac{N}{2}\log\{1+snr\}\nonumber\\
 &+&N\frac{\Gamma(N)}{\Gamma(N/2)^2}\sum_{m=1}^M
 \sum_{k=0}^{K}(-1)^{m+k}\frac{\Gamma(m+k)}{\Gamma(m)}\frac{snr^{m+k}}{k!m}A_m(N,1-k).
 \label{exp_final}
 }


\end{document}